\documentclass[10pt]{revtex4}
\usepackage{amssymb}
\usepackage{latexsym}
\usepackage{epsfig}
\begin{document}
\title{Thermodynamical description of modified generalized Chaplygin gas model of dark energy}
\author{H. Ebadi$^{1,2}$\footnote{hosseinebadi@tabrizu.ac.ir}, H. Moradpour$^2$\footnote{h.moradpour@riaam.ac.ir}}
\address{ $^1$ Astrophysics Department, Physics Faculty, University of Tabriz, Tabriz, Iran,\\
$^2$ Research Institute for Astronomy and Astrophysics of Maragha (RIAAM), P.O. Box 55134-441, Maragha, Iran.}

\begin{abstract}
We consider a universe filled by a modified generalized Chaplygin
gas together with a pressureless dark matter component. We get a
thermodynamical interpretation for the modified generalized
Chaplygin gas confined to the apparent horizon of FRW universe,
whiles dark sectors do not interact with each other. Thereinafter,
by taking into account a mutual interaction between the dark sectors
of the cosmos, we find a thermodynamical interpretation for
interacting modified generalized Chaplygin gas. Additionally,
probable relation between the thermal fluctuations of the system and
the assumed mutual interaction is investigated. Finally, we show
that if one wants to solve the coincidence problem by using this
mutual interaction, then the coupling constants of the interaction
will be constrained. The corresponding constraint is also addressed.
Moreover, the thermodynamic interpretation of using either a
generalized Chaplygin gas or a Chaplygin gas to describe dark energy
is also addressed throughout the paper.
\end{abstract}

\maketitle

\section{Introduction}
Recent observations imply an expanding universe whiles its rate of
expansion is extremely increased \cite{Rie,Rie1,Rie2,Rie3}. A
primary model introduced to explain this kind of expansion uses
Cosmological Constant (CC) in order to model the source of this
expansion \cite{roos}. In fact, Standard cosmology, including a
universe filled by a CC along with the cold dark matter (CDM) and
baryonic matters, is in line with the observations and standard
particle physics theory \cite{roos,c1,c2,c3,c4,c5}. In addition,
this model ($\Lambda CDM$) is satisfying the generalized second law
of thermodynamics in its current stage of expansion indicating that
the universe maintains this phase \cite{pavon2}. However, the
unknown nature of dominated fluid or CC along with the coincidence
and fine tuning problems are some of the unsatisfactory parts of
this model \cite{roos}.

Moreover, CC is classified into a more general class of fluids named
dark energy (DE) using to describe the current phase of expansion
\cite{ch14}. There are also numerous models introducing a dynamic DE
\cite{rev1,rev2,LBSV,pol,de1,de2,de3,de4,de5,de6,ven,GGDE,jcap,sc1,sc2,ph1,ph2,qu1,qu2,ke1,ke2,
ta1,ta2,di1,di2,set,set1,set2,set3,cw,cw1}. Additionally, It seems
that the DE models in which state parameter is not constant have a
better fitting with observations compared with CC
\cite{bet,bet1,bet2,bet3,bet4,bet5}. One of the models which
includes non-constant state parameter is Chaplygin gas which has
interesting features \cite{ch14,ch}. In this model, the pressure of
DE is related to its density as
\begin{eqnarray}
p_D=-\frac{B}{\rho_D},
\end{eqnarray}
where $B$ is a constant and subscript $\textmd{D}$ is used to
indicate that it is a model for DE. This model attracted more
investigators to itself
\cite{ch,ch1,ch2,ch3,ch4,ch5,ch6,ch7,ch8,ch9}. In addition, the
generalization of Chaplygin gas (GCG) is
\begin{eqnarray}\label{gcgd1}
p_D=-\frac{B}{\rho_D^{\alpha}},
\end{eqnarray}
whiles $0\leq \alpha\leq1$ and covers CG for $\alpha=1$ \cite{gch}.
On one hand, this model can be reinterpreted as an entangled mixture
of DE and DM \cite{int3}, while on the other hand, this dual role is
eliminated by observations \cite{int3,gch2}. Finally, we should note
that one can consider GCG as a candidate for dynamical DE
\cite{gch3,gch4}. It is also useful to mention that this model
preserves the thermodynamical equilibrium conditions \cite{pavon2}.
Moreover, this generalization can be modified (MGCG) as
\begin{eqnarray}\label{gcgd}
p_D=A\rho_D-\frac{B}{\rho_D^{\alpha}},
\end{eqnarray}
whiles we have again $0\leq \alpha\leq1$, whenever $A$ is a constant
\cite{ch6,gch}. Additionally, the pressure profiles of GCG and CG
are obtainable by substituting $A=0$ and $A=0$ together with
$\alpha=1$ respectively. It is easy to show that
\begin{eqnarray}\label{hdot1}
\rho_D=(\frac{B}{1+A}+\frac{C}{a^{3(1+A)(1+\alpha)}})^{\frac{1}{1+\alpha}},
\end{eqnarray}
as the density profile of MGCG while $C$ is an integration constant
and $A\neq-1$. MGCG also covers the radiation density profile for
$A=\frac{1}{3}$. Moreover, for $A=0$, it behaves as a pressureless
matter in early universe and blurts a cosmological constant-like
behavior in the last stage of the universe \cite{ch6,gch}. For the
state parameter $\omega_D=\frac{p_D}{\rho_D}$ we get
\begin{eqnarray}\label{omega}
\omega_D=A-B\frac{a^{3(1+A)(1+\alpha)}(1+A)}{Ba^{3(1+A)(1+\alpha)}+C(A+1)},
\end{eqnarray}
meaning that the MGCG model looks like a mixture of a prefect fluid
with $\omega_{pf}=A$ and a GCG with
$\omega_{GCG}=B\frac{a^{3(1+A)(1+\alpha)}(1+A)}{Ba^{3(1+A)(1+\alpha)}+C(A+1)}$
which is in line with Eq.~(\ref{gcgd}). We should note again that
the density profile and the state parameter of considering GCG are
obtainable by inserting $A=0$ in Eqs.~(\ref{hdot1})
and~(\ref{omega}) respectively. The result of considering CG are
also covered by substituting $A=0$ together with $\alpha=1$ in
Eqs.~(\ref{hdot1}) and~(\ref{omega}). Additionally,
$\omega_D\rightarrow -1$ in the long run limit ($a\gg1$) when
$A>-1$, and the state parameter of phantom regime is also accessible
in this limit for $A<-1$. Various aspects of MGCG, as a dynamical
model for DE, was also investigated
\cite{MGCG1,MGCG2,MGCG3,MGCG4,MGCG5,MGCG6,MGCG7}. Meanwhile, some
thermodynamical aspects of the Chaplygin gas models, including CG,
GCG and MGCG, can also be found in Refs.
\cite{ter1,ter2,ter3,ter4,ter5,ter6,ch11,ch11b,ch12,ch13}. The state
parameter, heat capacity in constant volume and thermodynamic
behavior of MGCG in terms of temperature and volume are studied in
\cite{ch12}. The mutual correspondence between MGCG and DBI-essence
as well as the validity of generalized second law of thermodynamics
in a FRW universe filled by a MGCG are shown in ref.~\cite{ch13}.
Moreover, it is also shown that the generalized second law of
thermodynamics may be valid on Hubble, particle and apparent horizon
of the FRW universe filled by a MGCG together with a non-interacting
magnetic field \cite{ch11}. A comprehensive review on the DE and
modified gravity which is also including the Chaplygin gas models
can be found in \cite{ch14}.

From observational point of view, the dark sectors of universe,
including DE and DM, can interact with each other
\cite{ch14,gdo1,gdo2,ob1,ob2,ob3,ob4,ob5,ob6,ob7}. Additionally,
such interactions may solve the coincidence problem
\cite{ob7,co1,co2,co3,co4,co5,co6,pavonz,wangpavon}, if they lead to
decay DE into DM \cite{pavonz,wangpavon}. These results are also in
line with the observations supporting the idea of using a dynamical
model to describe DE \cite{bet,bet1,bet2,bet3,bet4,bet5}. Such
likely mutual interactions between the Chaplygin gas models and DM
were also investigated
\cite{z1,z2,int,int1,int2,int3,co1,int5,pavonz,int7,int8,shs0}.
Thermodynamics of the GCG model interacting with CDM was studied
which shows that the second law of thermodynamics can be valid in
this scenario \cite{ch10}. Moreover, it seems that thermal
fluctuations may correct the entropy of gravitational systems
\cite{das}. These corrections are not limited to the gravitational
system and indeed, they are available in all thermodynamical systems
\cite{das}. In the cosmological setups, thermal fluctuations of
system may be attributed to the mutual interaction between the dark
sectors \cite{wangpavon,shs00,shs0,shs,shs1,hsm}. Therefore, it is
useful to study the thermodynamics of mutual interaction between the
dark sectors of cosmos helping us to get a more actual model for
describing DE.

Our goal in this paper is finding a thermodynamical interpretation
for the mutual interaction between MGCG, as a dynamical model for
DE, and DM by using thermal fluctuations approach
\cite{wangpavon,shs00,shs0,shs,shs1,hsm}. For this propose, we take
into account that our universe, described by the FRW metric
\begin{eqnarray}\label{frw}
ds^{2}=dt^{2}-a^{2}\left( t\right) \left[ \frac{dr^{2}}{1-kr^{2}}%
+r^{2}d\Omega ^{2}\right],
\end{eqnarray}
includes a MGCG and a pressureless DM component. For the FRW metric,
$a(t)$ is the scale factor while $k=-1,0,1$ is the curvature
parameter corresponding to open, flat and closed universes
respectively \cite{roos}. Because apparent horizon of dynamical
spacetimes can be considered as a causal boundary
\cite{Bak,Hay2,Hay22}, we use the corresponding horizon for the FRW
metric located at \cite{sheyw1,sheyw2}
\begin{eqnarray}\label{ah}
\tilde{r}_A=\frac{1}{\sqrt{H^2+\frac{k}{a(t)^2}}}.
\end{eqnarray}
Finally, since WMAP data indicates a flat universe, we set $k=0$ in
our study \cite{roos}.

The paper is organized as follows. In the next section, we consider
the FRW universe filled by a MGCG together with a pressureless DM
whiles they do not interact with each other. Thereinafter, we find a
relation for the entropy changes of MGCG. In section $\textmd{III}$,
the mutual interaction between the dark sectors is taken into
account. We will find thermodynamical description for the
interacting MGCG model. Throughout this paper, we address the
results of considering either GCG or CG instead of MGCG by using
their relation to MGCG. Section $\textmd{IV}$ is devoted for a
summary and concluding remarks. For the sake of simplicity, we take
$G=\hbar=c=1$ throughout this paper.

\section{Thermodynamical description of non-interacting modified generalization of Chaplygin gas}\label{VDE}
Friedmann equations for the FRW universe filled by a DE together
with a pressureless DM are
\begin{eqnarray}\label{fried1}
H^{2}=\frac{8\pi}{3}(\rho_{m}+\rho_{D}),
\end{eqnarray}
and
\begin{eqnarray}\label{fried2}
-2\frac{\ddot{a}}{a}-H^2=8\pi p_D,
\end{eqnarray}
where dot denotes derivative with respect to time, and
$H\equiv\frac{\dot{a}}{a}$ is the Hubble parameter. Moreover,
$\rho_i$ and $p_D$ are the density of i$^{\textmd{th}}$ fluid and
the pressure of DE respectively. If we define
$\rho_c\equiv3\frac{H^2}{8\pi}$ and use Eqs.~(\ref{fried1}), we get
\begin{eqnarray}\label{friedman11}
1=\Omega_D+\Omega_m,
\end{eqnarray}
where $\Omega_i=\frac{\rho_i}{\rho_c}$ is the fractional energy
density of the i$^{\textmd{th}}$ component of the dark sector.
Energy-momentum conservation law implies
\begin{eqnarray}\label{emc}
\dot{\rho}_{m}+\dot{\rho}_{D}+3H(\rho_{m}+p_m+\rho_D+p_D)=0.
\end{eqnarray}
where dot denotes again derivative with respect to time. The
energy-momentum conservation law can be decomposed into
\begin{eqnarray}\label{dmc1}
\dot{\rho}_{m}+3H\rho_{m}=0,
\end{eqnarray}
and
\begin{eqnarray}\label{dec1}
\dot{\rho}_{D}+3H(\rho_{D}+p_D)=0,
\end{eqnarray}
when the dark sectors do not interact with each other. For MGCG
confined to the apparent horizon of the flat FRW universe, the
Gibb's law implies
\begin{eqnarray}\label{flt}
TdS_{D}=dE_{D}+p_{D}dV.
\end{eqnarray}
In this equation, $S_D$ is the entropy of MGCG while $V=\frac{4\pi
}{3}\tilde{r}_{A}^{3}$ and $E_D=\rho _{D}V$ are the volume of the
flat FRW universe and the energy of MGCG respectively. Since
thermodynamic equilibrium condition implies that $T$ (the
temperature of MGCG) should be equal to that of the apparent horizon
\cite{Wanggg,sheyw2,Cai2}, we obtain
\begin{eqnarray}\label{temp}
T=\frac{H}{2\pi}=\frac{1}{2\pi \tilde{r}_{A}},
\end{eqnarray}
where we have used Eq.~(\ref{ah}) for the flat FRW universe ($k=0$).
Because
\begin{eqnarray}\label{vo}
dV=4\pi(\tilde{r}_{A})^{2}d\tilde{r}_{A}=-4\pi H^{-4}dH,
\end{eqnarray}
and
\begin{eqnarray}\label{ener}
dE_{D}=\rho_D dV+Vd\rho_D,
\end{eqnarray}
we get
\begin{eqnarray}\label{f1}
dS_D=\frac{2\pi}{H}((\rho_D+p_D)dV+Vd\rho),
\end{eqnarray}
leading to
\begin{eqnarray}\label{fl2}
dS^0_D=\frac{2\pi}{H_0^3}(1+A-\frac{B}{(\rho^0_D)^{\alpha+1}})(\frac{1}{(1+A-\frac{B}{(\rho^0_D)^{\alpha+1}})
+u_0}-\frac{3\Omega^0_D}{2})dH_0
\end{eqnarray}
for the entropy changes of the MGCG model. The subscript/superscript
($0$) is used to emphasize the non-interacting parameters, and we
have also defined
$u\equiv\frac{\Omega_m}{\Omega_D}=\frac{\rho_m}{\rho_D}$ for
simplicity. In deriving the above equation, we used
\begin{eqnarray}\label{rho11}
d\rho_D=\frac{\dot{\rho}_D}{\dot{H}}dH,
\end{eqnarray}
where $\dot{\rho}_D$ is evaluated by using Eq.~(\ref{dec1}). In
addition, one should take derivative with respect to time
from~(\ref{fried1}) and uses Eqs.~(\ref{gcgd}),~(\ref{fried2})
and~(\ref{emc}) to obtain the Raychaudhuri equation
\begin{eqnarray}\label{h}
\dot{H}_0=-4\pi \rho^0_D((1+A-\frac{B}{(\rho^0_D)^{\alpha+1}})+u_0).
\end{eqnarray}
One may use~(\ref{temp}) to get
\begin{eqnarray}\label{fl3}
dS^0_D=\frac{1}{2\pi
T_0^3}(1+A-\frac{B}{(\rho^0_D)^{\alpha+1}})(\frac{1}{(1+A-\frac{B}{(\rho^0_D)^{\alpha+1}})+\frac{1-\Omega_D^0}{\Omega_D^0}}
-\frac{3\Omega^0_D}{2})dT_0,
\end{eqnarray}
where we have used $u_0=\frac{1-\Omega_D^0}{\Omega_D^0}$ in getting
this equation. We should note again that the subscript/superscript
($0$) indicates that the calculations are done for the case in which
the dark sectors do not interact with each other. It is also useful
to mention that one can reach the results of considering the GCG and
prefect fluid models by inserting $A=0$ and $B=0$ in the above
relations respectively. Moreover, by inserting $A=0$ and $\alpha=1$
we will find the changes of the CG entropy. Briefly, we get a
thermodynamical description for the prefect fluid and CG models
confined to the apparent horizon, when the DE candidate does not
interact with the pressureless DM.
\section{Thermodynamic description of interacting MGCG}\label{Int}
For the case in which MGCG interacts with the pressureless DM,
decomposition of the energy-momentum conservation is read as
\begin{eqnarray}\label{emcigde}
\dot{\rho}_{m}+3H\rho_{m}=Q,
\end{eqnarray}
and
\begin{eqnarray}\label{emcigde2}
\dot{\rho}_{D}+3H\rho_{D}(1+\omega_{D})=-Q,
\end{eqnarray}
indicating that the energy will be transferred to DM from MGCG for
$Q>0$. Indeed, in order to solve the coincidence problem, DE should
slowly decay into DM. Therefore, the mutual interaction should be
very weak in accordance with the long life of the universe. Such
interaction terms have been discussed from phenomenological point of
view \cite{z1,z2}. The general interaction term is written as
$Q=3H(c_1\rho_m+c_2\rho_D)$ introduced in ref.~\cite{int} for the
first time, where $c_1$ and $c_2$ are the coupling constants. We
should note that this interaction term is the general form of those
introduced in refs.~\cite{int,int5,int7,int8,shs0}, and attracted
more attempts to itself \cite{int1,int2,int3,ch14}. Use
Eqs.~(\ref{fried1}),~(\ref{gcgd}),~(\ref{emcigde})
and~(\ref{emcigde2}) to get the Raychaudhuri equation as
\begin{eqnarray}\label{rech}
\dot{H}=-4\pi\rho_D((1+A-\frac{B}{(\rho_D)^{\alpha+1}})+u).
\end{eqnarray}
Bearing the definition of $u$ in mind and use~(\ref{friedman11}) to
get
\begin{eqnarray}\label{rho12}
d\rho_D=\frac{\dot{\rho}_D}{\dot{H}}dH=\frac{3H(\rho_D+p_D)+Q}{4\pi\rho_D((1+A-\frac{B}{(\rho_D)^{\alpha+1}})+
\frac{1-\Omega_D}{\Omega_D})},
\end{eqnarray}
where we have used~(\ref{emcigde2}) and~(\ref{gcgd}) to obtain the
last equation. Finally, by following the recipe of previous section,
we get
\begin{eqnarray}\label{entf}
\frac{dS_D}{dH}=\frac{2\pi}{H^3}(\frac{3H\rho_D(1+A-\frac{B}
{(\rho_D)^{\alpha+1}})+Q}{3H\rho_D((1+A-\frac{B}
{(\rho_D)^{\alpha+1}})+\frac{1-\Omega_D}{\Omega_D})}-\frac{3}{2}\Omega_D),
\end{eqnarray}
for the entropy changes of the interacting MGCG model. Indeed, this
mutual interaction affects the entropy changes of MGCG studied in
the previous section. It is argued that the horizon entropy has a
logarithmic correction which is due to the thermal fluctuations
\cite{das}. This modification of entropy is available in all
thermodynamical systems, and it is not limited to the gravitational
systems \cite{das}. In cosmological setups, it seems that the mutual
interaction between the dark sectors of the cosmos may induce week
thermal fluctuations to the cosmos contents leading to correct the
entropy \cite{wangpavon,shs00,shs0,shs,shs1,hsm}. In order to find a
relation between thermal fluctuations and the dark sectors mutual
interaction, we express the entropy of the interacting MGCG model as
\cite{das,wangpavon,shs00,shs0,shs,shs1,hsm}
\begin{eqnarray}\label{totalentropy}
S_D=S_D^0+S_D^1+S^2_D.
\end{eqnarray}
In the above equation, $S_D^0$ and $S_D^1=-\frac{1}{2}\ln CT_0^2$
are the entropy of MGCG, when there is no mutual interaction between
the dark sectors of the universe, and logarithmic correction to the
entropy, which is due to the thermal fluctuations, respectively.
$S^2_D$ also concerns higher order terms which are so week in the
gravitational and cosmological systems \cite{das,wangpavon}.
Moreover, $C=T_0\frac{dS_D^0}{dT_0}$ is the dimensionless heat
capacity, and it is also useful to mention again that this analysis
is valid for all thermodynamical systems \cite{das}. It is a matter
of calculation to show that
\begin{eqnarray}\label{C1}
C=T_0\frac{dS_D^0}{dT_0}=\frac{1}{2\pi
T_0^2}(1+A-\frac{B}{(\rho^0_D)^{\alpha+1}})(\frac{1}{(1+A-\frac{B}{(\rho^0_D)^{\alpha+1}})+\frac{1-\Omega^0_D}{\Omega^0_D}}
-\frac{3\Omega^0_D}{2}),
\end{eqnarray}
where we have used~(\ref{fl3}) to derive the above equation.
Combination Eqs.~(\ref{dec1}),~(\ref{rho11}) and~(\ref{h}) leads to
\begin{eqnarray}\label{rhod}
\frac{d\rho^0_D}{dH_0}=\frac{3H_0(1+A-\frac{B}{(\rho^0_D)^{\alpha+1}})}{4\pi((1+A-\frac{B}{\rho^0_D)^{\alpha+1}}+
\frac{1-\Omega^0_D}{\Omega^0_D})}
\end{eqnarray}
Therefore, we get
\begin{eqnarray}\label{entropys1}
\frac{dS_D^1}{dH_0}=-\frac{1}{2}[\frac{(1+A+\alpha\frac{B}
{(\rho^0_D)^{\alpha+1}})}{(\rho^0_D+A\rho^0_D-\frac{B}{(\rho^0_D)^{\alpha}})}\frac{d\rho^0_D}{dH_0}+
\frac{8\frac{\pi}{H^3}+\frac{-\frac{3H}{4\pi}+\frac{d\rho^0_D}{dH_0}(A+
\alpha \frac{B}{(\rho^0_D)^{\alpha+1}})}
{(A\rho^0_D-\frac{B}{(\rho^0_D)^{\alpha}}+\frac{3H^2}{8\pi})^2}}
{\frac{1}{A\rho^0_D-\frac{B}{(\rho^0_D)^{\alpha}}+\frac{3H^2}{8\pi}}-\frac{4\pi}{H^2}}],
\end{eqnarray}
where we have used
\begin{eqnarray}\label{entropys11}
\frac{d}{dH_0}=\frac{d\rho^0_D}{dH_0}\frac{d}{d\rho^0_D},
\end{eqnarray}
whiles $\frac{d\rho^0_D}{dH_0}$ can be found in Eq.~(\ref{rhod}).
From Eq.~(\ref{totalentropy}), we get
\begin{eqnarray}\label{drho2}
\frac{dS_D}{dH}=(\frac{dS_D^0}{dH_0}+\frac{dS_D^1}{dH_0}+\frac{dS_D^2}{dH_0})\frac{dH_0}{dH},
\end{eqnarray}
leading to
\begin{eqnarray}\label{drho23}
\frac{dS_D^2}{dH_0}=\frac{dS_D}{dH}\frac{dH}{dH_0}-\frac{dS_D^0}{dH_0}-\frac{dS_D^1}{dH_0},
\end{eqnarray}
whenever
\begin{eqnarray}
\frac{dH}{dH_0}=\frac{\dot{H}}{\dot{H}_0}=\frac{\rho_D((1+A-\frac{B}{(\rho_D)^{\alpha+1}})+\frac{1-\Omega_D}{\Omega_D})}
{\rho^0_D((1+A-\frac{B}{(\rho^0_D)^{\alpha+1}})+\frac{1-\Omega^0_D}{\Omega^0_D})}.
\end{eqnarray}
In deriving the last equation, we used Eqs.~(\ref{h})
and~(\ref{rech}). Eq.~(\ref{drho23}) gives us an expression for the
higher order terms of the thermal fluctuations ($S^2_D$) which was
omitted in the previous works
\cite{das,wangpavon,shs00,shs0,shs,shs1,hsm}. Moreover, we should
note that since the $S^0_D$ and $S^1_D$ terms have the major
contributions in the expression~(\ref{drho2}), one can withdraw the
$S^2_D$ contribution, and writes
\cite{das,wangpavon,shs00,shs0,shs,shs1,hsm}
\begin{eqnarray}\label{drho2}
\frac{dS_D}{dH}=(\frac{dS_D^0}{dH_0}+\frac{dS_D^1}{dH_0})\frac{dH_0}{dH}.
\end{eqnarray}
Indeed, this relation helps us to find an expression between the
mutual interaction of the dark sectors and the thermal fluctuations
of the universe, up to the first order. Therefore, we provided a
thermodynamical interpretation for the mutual interaction between
the dark sectors of the universe. In order to solve the coincidence
problem, DE should decay into DM meaning that the mutual interaction
between the dark sectors must meet the $Q>0$ condition
\cite{wangpavon}. By applying this condition to a general
interaction $Q=3H(c_1\rho_m+c_2\rho_D)$ and using
Eq.~(\ref{friedman11}) we get
\begin{eqnarray}\label{coin}
\frac{c_1}{c_2}>-\frac{\Omega_D}{1-\Omega_D}=-\frac{1}{u},
\end{eqnarray}
where we have used this fact that $H\rho_c>0$. Therefore, when the
coupling constants ($c_1$ and $c_2$) meet this condition, the mutual
interaction between the dark sectors may solve the coincidence
problem. Additionally, our surveys help us to find the corresponding
relation between the thermal fluctuations of the system and mutual
interaction between the dark sectors. Finally, it is useful to note
that by substituting $A=0$ along with $\alpha=0$ and $A=0$ in the
above formulas one can find the results of considering the
interacting CG and GCG models respectively.

\section{Summary and concluding remarks}
Since the recent observations admit an interaction between the dark
sectors of the cosmos \cite{ob1,ob2,ob3,ob4,ob5,ob6,ob7}, the hopes
to resolve the coincidence problem, by using this interaction, are
increased \cite{ob7,co1,co2,co3,co4,co5,co6,pavonz,wangpavon}. In
addition, it seems that such interactions induce thermal
fluctuations into the system
\cite{wangpavon,shs00,shs0,shs,shs1,hsm}, and therefore one should
be able to find a mutual relation between the thermal fluctuations
and such interactions. Moreover, there are observation evidences
indicating that dynamical DE models have a better fitting with
observations \cite{bet,bet1,bet2,bet3,bet4,bet5}. All of these data
motivate us to study the thermodynamics of mutual interaction
between MGCG and DM. Here, we have considered MGCG as a dynamical
model for DE and studied thermodynamics of the non-interacting MGCG
model. Therefore, we got its corresponding thermodynamical
description. In continue, we have focused on the interacting MGCG
model and studied its thermodynamics. We saw that the entropy of the
interacting MGCG model differs from the non-interacting case.
Additionally, bearing the thermal fluctuations theory along with the
Raychaudhuri equation in mind, we could find an expression for the
fluctuations due to the mutual interaction between the dark sectors.
The relation of our results to those of the GCG and CG models was
also addressed by pointing to the corresponding limits. Finally, we
showed that the solving of the coincidence problem conditions that
the coupling constants of the mutual interaction between the dark
sectors are not independent of each other. Indeed, they should
satisfy relation~(\ref{coin}). Briefly, our studies show that one
can get an expression for the mutual interaction between the dark
sectors by investigating the thermal fluctuations of the cosmos
components. Loosely speaking, We could find a thermodynamical
interpretation for the Chaplygin gas models, including the CG, GCG
and MGCG models, and their probable interaction with the
pressureless DM. The evolution of these fluctuations due to the
universe expansion have a great importance, because it can help us
to estimate the future of the cosmos. This could be an interesting
problem for future works.



\end{document}